\documentclass[reprint, amsmath, amssymb, aps, superscriptaddress, prx, longbibliography]{revtex4-2}

\usepackage{graphicx}
\usepackage{epstopdf}
\usepackage{dcolumn}
\usepackage{bm}
\usepackage{color}
\usepackage[colorlinks=true, urlcolor=blue, linkcolor=blue, citecolor=blue]{hyperref}
\usepackage{natbib}
\usepackage{amsbsy}
\usepackage{lipsum}
\usepackage{verbatim} 

\usepackage[dvipsnames]{xcolor}
\usepackage{pst-all}
\usepackage{ulem}

\usepackage{lineno}

\begin{document}

\title{
Two-dimensional topological Anderson insulator in a HgTe-based semimetal 
}

\author{D.\,A.\,Khudaiberdiev}
\affiliation{Institute of Solid State Physics, Technische Universität Wien, 1040 Vienna, Austria}

\author{Z.\,D.\,Kvon}
\affiliation{Institute of Semiconductor Physics, 630090 Novosibirsk, Russia}
\affiliation{Novosibirsk State University, Novosibirsk 630090, Russia}

\author{M.\,S.\,Ryzhkov}
\affiliation{Institute of Solid State Physics, Technische Universität Wien, 1040 Vienna, Austria}

\author{D.\,A.\,Kozlov}
\affiliation{Experimental and Applied Physics, University of Regensburg, D-93040 Regensburg, Germany}

\author{N.\,N.\,Mikhailov}
\affiliation{Institute of Semiconductor Physics, 630090 Novosibirsk, Russia}

\author{A.\,Pimenov}
\affiliation{Institute of Solid State Physics, Technische Universität Wien, 1040 Vienna, Austria}

\date{\today}

\begin{abstract}
We report the experimental observation of Anderson localization in two-dimensional (2D) electrons and holes in the bulk of HgTe quantum wells with a semimetallic spectrum and under strong disorder. 
In contrast, the one-dimensional (1D) edge channels, arising from the spectrum's inversion, demonstrate remarkable robustness against disorder due to topological protection.  
Strong disorder induces a mobility gap in the bulk, enabling access to the 1D edge states and thereby realizing the two-dimensional topological Anderson insulator (TAI) state.
Nonlocal transport measurements confirm the emergence of topologically protected edge channels.
The TAI state appears to be very sensitive to an external magnetic field applied perpendicular to the sample.  
Firstly, a small magnetic field of 30\,mT breaks the topological protection of 1D edge channels, thus turning the system into an ordinary Anderson insulator. Secondly, the magnetic field of 0.5\,T delocalizes 2D bulk electrons, transforming the system into a quantum Hall liquid.  
\end{abstract}

\maketitle

\section{Introduction}
    
Topological insulators (TI) represent a special state of condensed matter with an insulating bulk and conducting gapless states on the surface or the edge \cite{Hasan2010,Qi2011,Qi2010,Moore2007,Moore2010}. 
The existence of such materials is justified within a concept of topological ordering, which can be expressed in the form of invariants over the momentum space. 
In the presence of time-reversal symmetry, materials with band gaps (band insulators) are classified by the topological invariant $Z_2$ \cite{Kane2005}, which in the two-dimensional (2D) case can take two values, "1" or "0", thereby providing a distinction between topological and normal insulators. 
The 2D\,TI state is sometimes referred to as the
quantum spin Hall (QSH) effect\,\cite{Bernevig2006}. In contrast to the quantum Hall effect, the formation of the edge states in the QSH effect requires no magnetic field: the spin-up and spin-down electrons propagate along the edge in opposite directions with the time-reversal symmetry being conserved.
The first experimental evidence of the QSH effect was obtained in HgTe quantum wells (QWs) \cite{Konig2007} by demonstration of a resistance plateau around $h/2e^2$ in the longitudinal resistance of a mesoscopic Hall bar. 
This observation was further confirmed by nonlocal experiments in the ballistic \cite{Roth2009} and diffusive \cite{Gusev2011} transport regimes.

Just after the introduction of the TI state, the existence of topological Anderson insulators (TAI) was theoretically predicted\,\cite{Li2009}. 
The basic idea is that disorder or interaction could transform an ordinary insulator into a 2D\,TI. 
The first prediction inspired an intensive theoretical \cite{Jiang2009,Groth2009, Xing2011, Chen2017, Vu2022}  research expanding the first ideas into such areas as photonic crystals\,\cite{Stutzer2018, Liu2020, Ren2024}, cold atoms\,\cite{Meier2018,Nakajima2021}, electronic
circuits \cite{Zhang2019, Zhang2021}, or mechanics \cite{Shi2021}. 
In spite of these efforts, up to now there have been no experimental confirmation of the original idea of an electronic 2D\,TAI.

In this work we report an experimental observation of a two-dimensional topological Anderson insulator (2D\,TAI) in a disordered 14\,nm HgTe quantum well with a semimetallic spectrum.
In  HgTe based semimetals the 2D bulk conduction and valence bands overlap ($\sim 5$\,meV for 14\,nm QW \cite{Gospodaric2021,Vasilev2021,Khudaiberdiev2023}). Therefore, it is usually considered impossible to observe the 2D\,TI state for these systems, since the bulk states would always shunt the edge.
However, the recent observation of Anderson localization for both electrons and holes in this system due to strong disorder \cite{Kvon2021} opens new possibilities.
If the edge states arising from the inverted band structure are topologically protected from Anderson localization, it may be possible to achieve the 2D\,TI state by increasing disorder. This would localize the 2D bulk states, thus realizing a 2D\,TAI.
It is important to state a fundamental difference to the initial idea of the 2D\,TAI\,\cite{Li2009}, where the system transforms from an ordinary insulator to the 2D\,TI as a result of increasing disorder. 
On the contrary, in present work the bulk state is initially conducting and becomes insulating due to opening of the mobility gap in the disordered state. In our case the edge states in the clean regime are shunted by the conducting bulk and only start to be observed in the gapped regime due to disorder.
Although in both cases disorder is crucial to the emergence of the 2D\,TI state, we believe that the present approach is even closer to the original idea of P.\,W.\,Anderson \cite{Anderson1958}. 

The investigated samples are Hall-like meso-structures fabricated on the basis of strongly disordered 14\,nm HgTe (013) QWs grown by molecular-beam epitaxy. 
A dielectric layer (200 nm of SiO$_2$) was deposited on the sample and then covered by a TiAu gate. 
The schematic cross section of the QWs is shown in Fig.\,\ref{Figure1}(a), and the sketch of the Hall bar with the gate is given in Fig.\,\ref{Figure1}(b). 
The measurements were carried out using the standard phase-detecting technique at temperatures of 80\,mK--10\,K and in magnetic fields up to 1.5\,T. 
The oscillator frequency was in the range 0.3--6\,Hz and the driving currents were 1\,pA–-10\,nA depending on the measurement conditions.

\section{Semimetallic state}

\begin{figure}
        \includegraphics[width=0.48\textwidth]{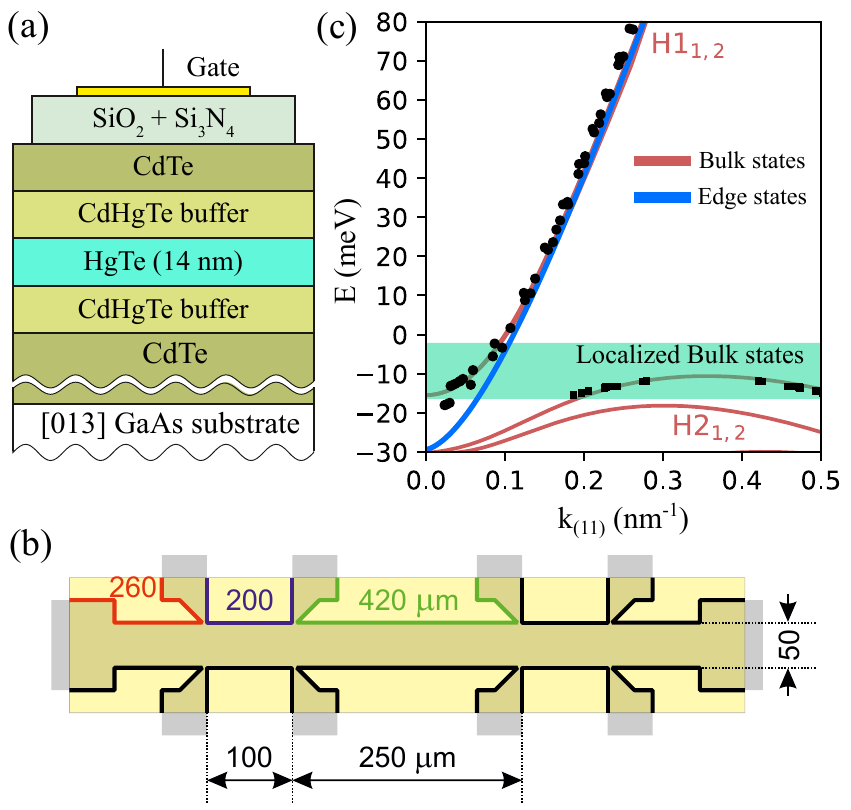}
        \caption{\label{Figure1}
        (a) Cross-section of the studied structure. 
        (b) Schematic top view of the Hall bars with the gate (indicated by dark yellow).
        (c) Energy spectrum of the studied system adapted from Ref.\,\cite{Gospodaric2021}. Red lines are the $\boldsymbol{k \cdot p}$ calculations of the bulk 2D spectrum, with the black dots as experimentally reconstructed spectrum by the cyclotron-resonance measurements. The turquoise strip represents a mobility gap of the 2D bulk spectrum caused by the Anderson localization. Blue line is an expected topologically protected 1D edge channel spectrum. 
        }
    \end{figure}

The energy spectrum of such QWs was reconstructed by the THz spectroscopy technique\,\cite{Gospodaric2021}, which agreed with the $\boldsymbol{k \cdot p}$ calculated spectrum. 
The spectrum from Ref.~\cite{Gospodaric2021} is adapted to Fig.\,\ref{Figure1}(c).
According to these results, the 14\,nm HgTe quantum well is a semimetal with conduction and valence bands overlap of about 5\,meV.

The system exhibits either purely electronic or semimetallic states depending on the gate voltage ($V_g$). 
The gate voltage dependencies of electron ($N_s$) and hole ($P_s$) densities at the temperature of $T=4$\,K, determined through Drude fitting of magnetotransport measurements, are shown in Fig.\,\ref{Figure2}(e).
An example of semimetallic magnetotransport behavior is provided in the inset.
The difference $N_s-P_s$ varies linearly with the gate voltage, crossing zero at the charge neutrality point (CNP) with $V_g^{CNP} = -0.62$\,V.
Due to the semimetallic nature of the system, the charge densities at the CNP are nonzero, with $N_s = P_s \approx 2\times10^{10}$\,cm$^{-2}$, and to the left from the CNP the system is in the deep semimetallic regime with $N_s \ll P_s$.  
Notably, the semimetallic state persists down to the base temperature of $T \approx 80$\,mK for strongly negative gate voltages. 
This is evidenced by the sign-alternating Hall resistivity, as shown in Fig.\,\ref{Figure2}(f). These results demonstrate that the overlap between the conduction and valence bands is maintained across the entire temperature range, confirming the absence of a bandgap in the studied system.

\section{2D TAI in nonlocal transport}

   \begin{figure*}
        \includegraphics[width=0.99\textwidth]{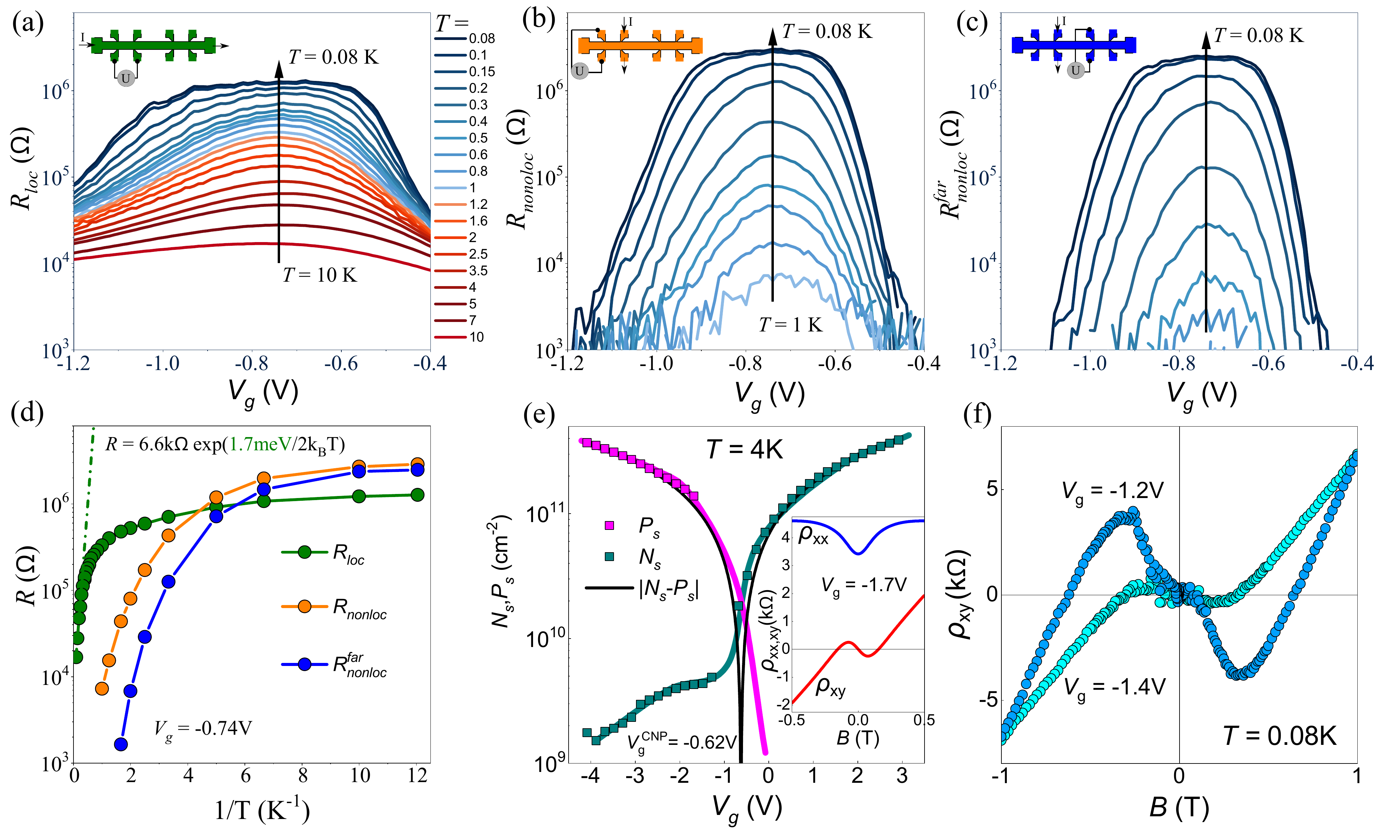}
        \caption{\label{Figure2}
        (a) Local, (b) nonlocal, and (c) far nonlocal resistances as functions of the gate voltage ($V_g$) at various temperatures (the legend is valid for all three graphs). (d) Local and both nonlocal resistances plotted as functions of the inverse temperature at $V_g = -0.74$\,V corresponding to the resistance maxima, as indicated by arrows in panels (a), (b), and (c).
        (e) Gate voltage dependence of the electron ($N_s$, pink squares) and hole ($P_s$, dark cyan squares) concentrations extracted using the two-component Drude model. The inset shows an example of the fitted data. Corresponding colored lines guide the eye. The black line represents the absolute difference $|N_s - P_s|$, highlighting the charge neutrality point ($V_g^{CNP} = -0.62$ V).
        (f) Magnetic field dependence of the Hall resistance $\rho_{xy}$ at the base temperature for $V_g = -1.2$\,V (dark blue) and $V_g = -1.4$\,V (cyan). 
        }
    \end{figure*}

The observation of both local ($R_{\mathrm{loc}}$) and nonlocal ($R_{\mathrm{nonloc}}$) resistances in the absence of magnetic field is generally considered \cite{Roth2009,Gusev2011} as a direct indication of a 2D\,TI state. 
In that case the current flows through the sample along its edges, and a measurable resistance, $R \propto V (I=const)$ exists regardless of the positions of the voltage probes with respect to the current contacts. 
Fig.\,\ref{Figure2} presents the results of the resistance measurements in local and nonlocal configurations. 
Figures \ref{Figure2}(a), \ref{Figure2}(b) and \ref{Figure2}(c) show the gate voltage dependencies of $R_{\mathrm{loc}}$, $R_{\mathrm{nonloc}}$ and $R_{\mathrm{nonloc}}^{\mathrm{far}}$ respectively. 
At a first glance, qualitatively similar behavior of local and nonlocal resistances is observed: 
both exhibit a broad maximum ($V_g^{max} = -0.74$\,V), slightly shifted to negative voltages from the CNP, followed by a sharp decrease on either side of the curve. 
The detailed analysis of the data reveals the fundamental difference between local and nonlocal resistances: firstly, the maximum plateau in $R_{\mathrm{loc}}(V_g)$ is notably wider than that in $R_{\mathrm{nonloc}}(V_g)$ and $R_{\mathrm{nonloc}}^{\mathrm{far}}(V_g)$, and, secondly, both nonlocal responses vanish for increasing temperatures while the local one remains finite up to the highest temperatures. 
Fig.\,\ref{Figure2}(d) shows all three resistances as the function of temperature at $V_g = V_g^{max}$. 
For high temperatures above $\sim 1$\,K, all resistances increase exponentially as the temperature decreases. The activation energy for the local resistance is 1.7\,meV (the exponential slope is indicated by dash-dotted green line). 
For temperatures $T > 0.4$\,K, the condition $R_{\mathrm{loc}} \gg R_{\mathrm{nonloc}}$, $R_{\mathrm{nonloc}}^{\mathrm{far}}$ holds,   indicating that the transport mechanism is dominated by the bulk.
At lower temperatures $T \lesssim 0.5$\,K, the resistance increase slows down and eventually saturates. In this regime both $R_{\mathrm{nonloc}}$ and $R_{\mathrm{nonloc}}^{\mathrm{far}}$ become comparable to or even exceed $R_{\mathrm{loc}}$, reflecting a transition to a transport mechanism dominated by edge contributions.

\begin{figure}   
    \includegraphics[width=0.495\textwidth]{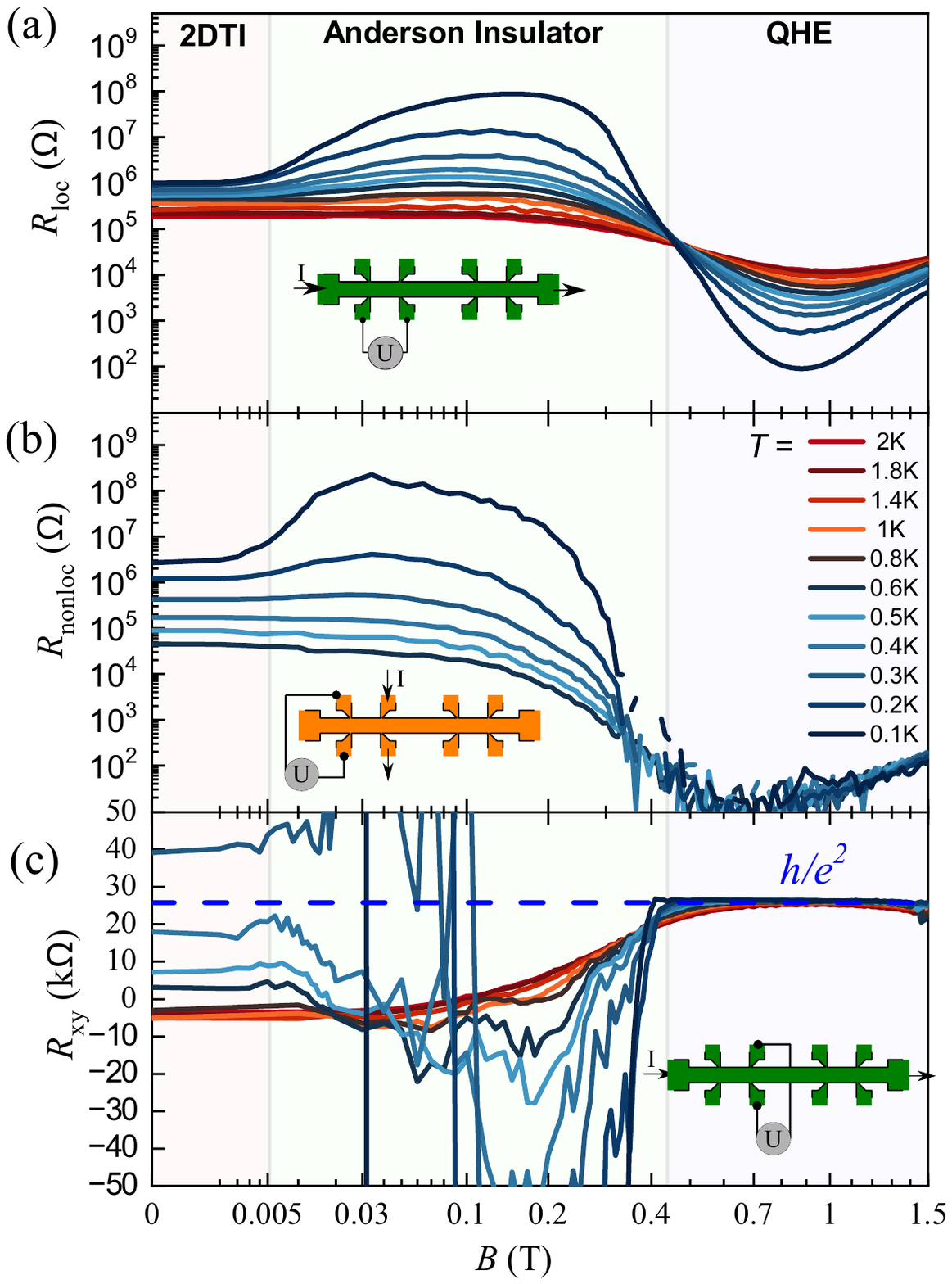}
    \caption{\label{Figure3}
    Magneto-resistance of a topological Anderson insulator. (a) Local, (b) nonlocal and (c) Hall resistances at  $V_g$ = -0.74\,V in the temperature range from 100\,mK to 2\,K.
    The magnetic field axis is scaled as the cube root to emphasize both 2D\,TI-to-Anderson insulator and Anderson insulator-to-quantum Hall liquid transitions.
    }
\end{figure}

The observed behavior of local and nonlocal responses is typical for the 2D\,TI, as it follows from the fundamental difference in the relative contributions of the bulk and edge conductivities for local and nonlocal configurations\,\cite{Roth2009,Olshanetsky2015,Kvon2020,Benlenqwanssa2024}. 
At high temperatures, the bulk contribution dominates, resulting in 
nonlocal resistances being exponentially smaller than the local one. As the temperature decreases, particularly in the presence of strong disorder, the bulk contribution to transport diminishes exponentially due to localization effects. Consequently, nonlocal transport along the sample edges becomes the dominant mechanism.
In this regime, the ratio of voltages or resistances measured in local and nonlocal configurations is governed by the currents flowing along the sample edges and the positioning of the voltage probes. In the extreme case where current flows exclusively along the sample edges, the geometry of the sample causes the voltage drop in the local configuration to become smaller than the one in the nonlocal configuration.
In extreme regime where the current flows solely along the sample edges, due to the geometry of the sample the voltage drop in the local configuration is even lower than the ones in nonlocal.
In addition,  for $T < 0.2$\,K and for all investigated configurations, the resistance maximum $R_{\mathrm{nonloc}} (V_g)$ is more than two orders of magnitude larger than $h/e^2$, thus indicating that the edge transport is diffusive with mean free path $l_{edge} \approx 1$\,$\mathrm{\mu}$m. This value is quite is typical for disordered 2D\,TI \cite{Konig2007, Gusev2011, Olshanetsky2015}. 
Although scattering occurs within the 1D edge channels, topological protection prevents the complete localization of these states, contrary to predictions from conventional scaling theory\,\cite{Abrahams1979}. The 1D conductivity scaling under topological protection is rather classical with only incoherent scattering. However, breaking time-reversal symmetry (see Fig.\,\ref{Figure3}) restores conventional localization mechanisms\,\cite{Piatrusha2019}.

The described relation between 2D bulk and 1D edge states of 2D\,TI with respect to Anderson localization represent the main result of the this work.
While 2D bulk electrons and holes are strongly localized with an exponential temperature dependence of  resistance, the 1D edge current states experience no Anderson localization due to  topological protection and continue to conduct even at the lowest temperatures.
This finding emphasizes an incredible robustness of the topologically protected edge channels, which survived under strong disorder even when the bulk states were localized.  
The fact that 2D\,TI state is only achievable due to Anderson localization indicates that we are dealing with a new kind of 2D\,TI -- 2D topological Anderson insulator with a 2D band of the localized states and 1D conducting states.




Further on, one might notice that in the vicinity of the CNP and for $T < 0.2$\,K, gate voltage dependencies of all resistances are rather weak (see Fig.\,\ref{Figure2}(a-c)).
In this regime where the edge transport dominates, the maxima are wide, such that the resistance does not change in the broad voltage range of around 0.3\,V. 
This behavior significantly differs from the one observed in narrow gap 2D\,TI \cite{Olshanetsky2015} where quite sharp resistance peak is observed, and both $R_{\mathrm{loc}}(V_g)$ and $R_{\mathrm{nonloc}}(V_g)$ are strongly asymmetrical with respect to the maximum position. 
We suggest the following explanation for this observation, taking into account the energy spectrum the 2D topological Anderson insulator given in Fig.\,\ref{Figure1}(c).
The conduction (H1) and valence (H2) bands (Fig.\,\ref{Figure1}(c)) in our system are mostly formed by the $\Gamma_8$ band, while the $\Gamma_6$ band state E1 is pushed deep into the energy spectrum\,\cite{Konig2008a}. This implies that the Dirac point of the edge states is also deep inside the valence band, thus in experimentally accessible gate  range the Fermi level remains always in the electron part of the edge spectrum. This regime roughly corresponds to the turquoise area in the band spectrum shown in Fig.\,\ref{Figure1}(c).
In the regime of localized bulk states and the edge spectrum far from the Dirac point, the maximum of resistivity in the edge transport regime is wide and nearly symmetric.



\section{Phase transitions under perpendicular magnetic field}

To further investigate the 2D\,TAI state we performed the magnetotransport measurements (magnetic field is perpendicular to the sample's plane) at the gate voltage $V_g = V_g^{max}$. 
The magnetic field dependencies of the local $R_{loc}(B)$, nonlocal $R_{nonloc}(B)$ and Hall $R_{xy}(B)$ resistances are shown in Fig.\,\ref{Figure3}.
We identify three transport regimes of the system depending on the applied magnetic field: 2D\,TI, ordinary Anderson insulator and the quantum Hall liquid.

At zero magnetic field the system is in the 2D\,TI state, as discussed above.
At low temperatures $R_{\mathrm{loc}} \ll R_{\mathrm{nonloc}}$, which indicates that the transport is going predominantly along the sample's edges. The low-temperature Hall resistance $R_{xy}$ takes random values due to possible asymmetry of the top and bottom paths of the electrons along the edges, but it is still by orders of magnitude smaller than both local and nonlocal resistances. 

Then, at low magnetic fields fields between 5\,mT and 200\,mT we observe the breakdown of 2D\,TI topological protection, which leads the system into the (ordinary) Anderson insulator regime with the mobility gap preventing the current flow.
Both, $R_{nonloc}$ and $R_{loc}$ grow by about two orders of magnitude exceeding the values of 100\,M$\Omega$. Compared to the field-free regime, the edge states are not topologically protected due to the breaking of time-reversal symmetry and the magnetic field-induced helical edge-states localization.
This effect is similar to the one observed in 8 nm HgTe QWs \cite{Piatrusha2019} without strong disorder.
Observed time-reversal symmetry breaking supports the 2D\,TAI picture.

Finally, in high magnetic fields, the system is in the quantum Hall liquid state. As one can see in both $R_{loc}$ and $R_{xy}$, there is a quantum phase transition from Anderson insulator to the quantum Hall liquid ~\cite{Jiang1993,Hughes1994,Shahar1995} at critical magnetic field of $B_c \approx 0.5$\,T and with critical resistance $\rho_{xx}(B_c) \approx 1.5$\,$h/e^2$.  It is worth noting that compared  to the results on  GaAs-based QWs~\cite{Jiang1993,Hughes1994}, the transition in our case occurs at much lower (almost one order of magnitude) magnetic fields and the filling factor $\nu$ is "1" rather than "2". 
It is not surprising because the 2D electrons in HgTe QWs have completely different Landau level spectrum with  larger cyclotron gap, large Zeeman splitting ($g \mu_B > \hbar\omega_c/2$) and because the two-component electron-hole system is affected by Anderson localization\,\cite{Kvon2021}. 
The holes are apparently still localized and do not contribute to the Hall conductivity allowing perfect quantization.

A transition from an ordinary 2D\,TI with the bulk gap to the quantum Hall state was predicted theoretically ~\cite{Scharf2012,Durnev2016}, and studied experimentally ~\cite{Olshanetsky2018, Shamim2022a}. 
It is necessary to note that in these works the bulk was characterized by a band gap contrasted to the mobility gap due to disorder in our case.

\section{Conclusion}
In this work we investigated a 14\,nm wide HgTe quantum well with a semimetallic energy spectrum. Our findings reveal that in the presence of disorder, the 2D bulk electrons and holes exhibit strong Anderson localization, but at the same time, the topologically protected 1D edge current states, emerged from the inversion of the spectrum, remain unaffected.
This underscores the robustness of the topologically protected 1D edge channels and their dominance over the 2D bulk states in stark contrast to the predictions of scaling theory, which suggests that 1D transport should be more vulnerable than 2D transport.
As a result, we demonstrate the realization of a new type of 2D topological insulator -- a 2D topological Anderson insulator with a localized bulk band.
Furthermore, in the presence of perpendicular magnetic field we observe two quantum phase transitions. 
An external field as small as 30\,mT is enough to break the topological protection and to localize the 1D edge channels, which leads to an ordinary Anderson insulator state. In higher fields above 0.5\,T we observe the transition from the Anderson insulator to the quantum Hall liquid.
These transitions raise a fundamental question of how the helical edge states in 2D\,TI are transformed into the chiral edge states of the QHE.

\textbf{Acknowledgements}\\
The financial support of this work by the Russian Science Foundation (Grant No. 23-72-30003 ) is
acknowledged.




\end{document}